\documentclass[ALICE,manyauthors]{cernphprep}
\usepackage{color} 
\usepackage{bm}
\usepackage{cite}

\newcommand{\jpsi}{\rm J/$\psi$}
\newcommand{\psip}{$\psi(\rm 2S)$}

\usepackage{rotating}
\usepackage{lineno}
\usepackage{hyperref}
\graphicspath{{./Figures/}{./}}

\begin{document}%
%%%%%%%%%%%%% ptdr definitions %%%%%%%%%%%%%%%%%%%%%
%
%%%%%%%%%%%%%%%  Title page %%%%%%%%%%%%%%%%%%%%%%%%
%
\begin{titlepage}
\PHyear{2016}      % required, obtained from PH
\PHnumber{046}
\PHdate{29 February}                 % required
%\EXPnumber{ALICE-INT-2010-9999}     % optional
%\EXPdate{12 October 2010}           % optional
%
%
%%% Put your own title + short title here:
\title{Centrality dependence of $\mathbf{\psi}$(2S) suppression in \mbox{p-Pb} collisions at $\mathbf{\sqrt{{\textit s}_{\rm NN}}}$~=~5.02 TeV}

\ShortTitle{Centrality dependence of $\mathbf{\psi}$(2S) suppression in \mbox{p-Pb} collisions}
   % appears on right page headers
%
%%% Do not change the next lines!
\Collaboration{ALICE Collaboration\thanks{See Appendix~\ref{app:collab} for the list of collaboration members}}
\ShortAuthor{ALICE Collaboration}      % appears on left page headers, do not change
\begin{abstract}
The inclusive production of the \psip\ charmonium state was studied as a function of centrality in \mbox{p-Pb} collisions at the nucleon-nucleon center of mass energy $\sqrt{s_{\rm NN}}$ = 5.02~TeV at the CERN LHC. The measurement was performed with the ALICE detector in the center of mass rapidity ranges 
$-4.46<y_{\rm cms}<-2.96$ and $2.03<y_{\rm cms}<3.53$, down to zero transverse momentum, by reconstructing the \psip\ decay to a muon pair.
The \psip\ production cross section $\sigma_{\psi(\rm 2S)}$ is presented as a function of the collision centrality, which is estimated through the energy deposited in forward rapidity calorimeters.
The relative strength of nuclear effects on the \psip\ and on the corresponding 1S charmonium state J/$\psi$ is then studied by means of the double ratio of cross sections $[\sigma_{\psi(\rm 2S)}/\sigma_{\rm J/\psi}]_{\rm pPb}/[\sigma_{\psi(\rm 2S)}/\sigma_{\rm J/\psi}]_{\rm pp}$ between \mbox{p-Pb} and \mbox{pp} collisions, and by the values of the nuclear modification factors for the two charmonium states. The results show a large suppression of \psip\ production relative to the \jpsi\ at backward (negative) rapidity, corresponding to the flight direction of the Pb-nucleus, while at forward (positive) rapidity the suppressions of the two states are comparable. 
Finally, comparisons to results from lower energy experiments and to available theoretical models are presented.

\end{abstract}
\end{titlepage}
\setcounter{page}{2}
%
%\linenumbers
\section{Introduction}

Charmonia are bound states of a charm and an anticharm quark ($c\overline c$), and represent an important testing ground for the properties of the strong interaction. In high-energy proton-proton collisions, the charmonium production process is usually factorized in two steps: the creation of a $c\overline c$ pair followed, on a longer time scale, by the binding and emission of one or more gluons that brings the pair to a colour singlet state. This process is described reasonably by theoretical models inspired by  Quantum Chromodynamics (QCD)~\cite{Brambilla:2010cs}, although a quantitative evaluation of the production cross sections and polarization of the charmonium states still meets difficulties~\cite{Brambilla:2010cs,Andronic:2015wma}.

If a charmonium state is produced within the nuclear medium, as can happen in  proton-nucleus collisions, several effects become important and might influence the charmonium formation. In particular, the modification in the nucleus of the parton distribution functions (shadowing/anti-shadowing)~\cite{Eskola:2009uj,deFlorian:2011fp,Hirai:2007sx}, can lead to a suppression or an enhancement of the charmonium production.  Furthermore, the incoming partons, as well as the outgoing $c\overline c$ pair, may lose energy in the nuclear medium, altering the differential distributions of the produced charmonium state~\cite{Arleo:2012hn}. Finally, once the bound state is formed, it may be dissociated via collisions within nuclear matter~\cite{Vogt:2001ky,Kopeliovich:1991pu,McGlinchey:2012bp}. However, the formation of the final-state resonance occurs in a finite time $\tau_{\rm f}$ which, depending on the kinematics of the $c\overline c$ pair and on the collision energy, may be longer than its 
crossing time, $\tau_{\rm c}$, in the nucleus. 

Among the narrow charmonium states, i.e. those with a mass smaller than twice the mass of the lightest D mesons, we address in this paper the vector states (J$^{\rm PC}$ = 1$^{\rm --}$) J/$\psi$, characterized by a binding energy $\Delta  E\sim 650$~MeV (corresponding to the mass gap to the open charm threshold), and the weakly bound \psip, with $\Delta  E\sim 50$~MeV~\cite{Satz:2005hx}.
%Several charmonium states exist, corresponding to various possible sets of quantum numbers for the $c\overline c$ pair. We restrict our considerations to the narrow states with a mass smaller than twice the mass of the lightest D mesons, and in particular to vector states (J$^{\rm PC}$ = 1$^{\rm --}$) which, decaying into lepton pairs, can be easily reconstructed from the experimental point of view. In this case, one is left with the most strongly bound \jpsi\ state, characterized by a binding energy $\Delta  E\sim 650$~MeV (corresponding to the mass gap to the open charm threshold), and the weakly bound \psip, with $\Delta  E\sim 50$~MeV~\cite{Satz:2005hx}.
A comparison of the production cross section of the two states in proton-nucleus collisions offers interesting insights into the size of the various cold nuclear matter (CNM) effects outlined above. In particular, shadowing acts on the initial state partons and has a nearly identical size for the two resonances~\cite{Ferreiro:2014bia,Ferreiro:2012mm}. Therefore, its effect largely cancels out when studying the ratio of their production cross sections.  Also coherent energy loss mechanisms~\cite{Arleo:2012hn}, have a similar effect on the two resonances, due to the fact that they act on a compact ${c\overline c}$ pair not yet evolved into a final color singlet state. 
On the contrary, the break-up probability of the final resonance inside the nucleus should be much larger for the weakly bound \psip~\cite{Vogt:1999dw}.

Early results on \jpsi\ and \psip\ production in proton-nucleus collisions were obtained at fixed target experiments by E866~\cite{Leitch:1999ea} at FNAL ($\sqrt{s_{\rm NN}} = 63$~GeV), by HERA-B~\cite{Abt:2006va} at HERA ($\sqrt{s_{\rm NN}} = 39$~GeV) and by NA38, NA50, NA60~\cite{Alessandro:2006jt,Alessandro:2003pc,Arnaldi:2010ky} at the CERN SPS ($\sqrt{s_{\rm NN}} = 17-29$~GeV). At mid-rapidity, i.e., close to $y_{\rm cms}=0$, the relative production cross section $\sigma_{\psi(\rm 2S)}/\sigma_{\rm J/\psi}$ was found to decrease rather strongly for increasing mass number of the nuclear target. Since part of the kinematic domain accessed at fixed target energies is characterized by $\tau_{\rm f}<\tau_{\rm c}$~\cite{McGlinchey:2012bp}, such an observation can indeed be related to a stronger break-up effect on the weakly bound \psip.

At collider energies, it becomes technically more difficult to have data samples corresponding to various nuclear colliding species. Therefore, in order to vary the thickness of CNM crossed by the $c\overline c$ pair, one can rather select classes of events based on estimators of the geometry (centrality) of the collision, corresponding to various ranges in the number of nucleon-nucleon collisions $N_{\rm coll}$. This procedure was followed by the PHENIX experiment at RHIC, which studied the nuclear modification factors, defined as the ratio between the measured yields in \mbox{d-Au} and proton-proton collisions, normalized to $N_{\rm coll}$, for the \jpsi\ and \psip\ resonances at 
mid-rapidity~\cite{Adare:2013ezl}. At $\sqrt{s_{\rm NN}} = 200$~GeV, the nuclear modification factors were smaller by a factor $\sim 3$ for \psip\ relative to \jpsi\ for  central events, indicating a stronger suppression for \psip. However, such an observation is surprising since for mid-rapidity production at RHIC energies the time spent by the $c\overline c$ pair in the nucleus ($\tau_{\rm c}<0.05$ fm/$c$) is below 
the formation time of the final-state resonance (most theory estimates~\cite{McGlinchey:2012bp,Hufner:2000jb,Kharzeev:1999bh} give $\tau_{\rm f}>0.15$ fm/$c$). In such a situation, one would rather expect a similar suppression for the \jpsi\ and \psip\ states.

At the LHC, centrality-integrated results on the \psip\ and \jpsi\ resonances for \mbox{p-Pb} collisions at $\sqrt{s_{\rm NN}}=5.02$~TeV were obtained by ALICE~\cite{Abelev:2014zpa,Abelev:2013yxa} and LHCb~\cite{Aaij:2016eyl,Aaij:2013zxa}. At both forward (positive) and backward (negative) rapidities, corresponding to the p-going and Pb-going directions respectively, a significantly larger suppression of \psip\ compared to \jpsi\ was observed, relative to proton-proton collisions. Again, this result was unexpected, as the $\tau_{\rm c}$ values are either at most the same order of magnitude (at negative $y_{\rm cms}$) or more than two orders of magnitude smaller (at positive $y_{\rm cms}$) than $\tau_{\rm f}$~\cite{Abelev:2014zpa}.
Therefore, additional effects, as the interaction of the loosely bound \psip\ with a hadronic or partonic medium produced in the collision, might be necessary in order to explain the results~\cite{Ferreiro:2014bia,Du:2015wha}.

As outlined above, a differential measurement as a function of the collisions centrality is equivalent to a study of the propagation of the $c\overline c$ pairs over various thicknesses of CNM. In this Letter, we go in that direction by showing results obtained by the ALICE Collaboration on \psip\ studies in \mbox{p-Pb} collisions as a function of centrality, estimated through the energy deposited at very forward rapidity by the remnants of the Pb-nucleus. 
The corresponding \jpsi\ studies were published in~\cite{Adam:2015jsa}.
In Sect.~2 we give a brief overview of the experimental apparatus and run conditions. Sect.~3 presents details on the analysis procedure, while Sect.~4 is dedicated to the results. The conclusions are presented in Sect.~5.

\section{Experimental conditions}

The analysis presented in this Letter is based on the detection of the $\psi(2S)\rightarrow\mu^+\mu^-$ decay in the forward muon  spectrometer of  ALICE, described in detail elsewhere~\cite{Aamodt:2011gj,Aamodt:2008zz}. This detector covers the pseudorapidity range $-4<\eta_{\rm lab}<-2.5$ and includes a 3~T$\cdot$ m dipole magnet and five stations of tracking chambers, the central one being inside the magnet gap. A main absorber (10 interaction lengths thick) is positioned between the ALICE interaction point and the tracking system, in order to remove hadrons. A second absorber is placed downstream of the tracking detectors. It removes the remaining hadrons and low-momentum muons produced predominantly from $\pi$ and $K$ decays, and is followed by two stations of trigger chambers that select muon candidates based on their transverse momentum ($p_{\rm T}$).
In addition to the muon spectrometer, the first two layers of the Inner Tracking System (SPD, i.e., Silicon Pixel Detectors, the first covering $|\eta_{\rm lab}|<2.0$ and the second $|\eta_{\rm lab}|<1.4$)~\cite{Aamodt:2010aa} are used for the determination of the position of the interaction vertex. The two V0 scintillator hodoscopes (covering  $-3.7<\eta_{\rm lab}<-1.7$ and $2.8<\eta_{\rm lab}<5.1$, respectively) are used for triggering purposes~\cite{Abbas:2013taa}. Finally, two sets of Zero-Degree Calorimeters (ZDC), positioned at 112.5~m on the two sides of the interaction point, each one including a neutron calorimeter (ZN) and a proton calorimeter (ZP), are used to clean-up the event sample from interactions occurring out of the nominal bunches and for the centrality estimate~\cite{ALICE:2012aa,Adam:2014qja}.

The data-taking conditions were described in ~\cite{Abelev:2013yxa,Abelev:2014ffa} and are briefly stated here. Two data samples were taken, corresponding to the p-beam or the Pb-beam going in the direction of the muon spectrometer, and labelled in the following as \mbox{p-Pb} and \mbox{Pb-p}, respectively. The integrated luminosities were $L_{\rm int}^{\rm pPb}=5.01\pm0.19$ nb$^{-1}$ and $L_{\rm int}^{\rm Pbp}=5.81\pm0.20$ nb$^{-1}$~\cite{Abelev:2014epa}.
The events used in this analysis were collected requiring a coincidence between a minimum bias (MB) trigger condition, defined by the logical AND of signals on the two V0 hodoscopes ($>$99\% efficiency for non-single diffractive events), and the detection of two candidate opposite-sign tracks in the trigger system of the muon spectrometer. A $p^{\mu}_{\rm T}>0.5$~GeV/$c$ cut on such tracks was also imposed at the trigger level. The offline event selection, the muon reconstruction and identification criteria and the kinematic and quality cuts applied at the single-muon and dimuon levels have already been described in Refs.~\cite{Abelev:2013yxa,Abelev:2014zpa,Adam:2015iga,Adam:2015jsa}. In particular, the covered dimuon rapidity ranges were $2.03<y_{\rm cms}<3.53$ and $-4.46<y_{\rm cms}<-2.96$ for the \mbox{p-Pb} and \mbox{Pb-p}  configurations, respectively. 

\section{Data analysis}
\label{sec:3}

In this Section, the evaluation of the various elements that enter the cross
section measurements and the nuclear modification factor calculations are 
described.

The centrality selection and the determination of $N_{\rm coll}$ are based on a hybrid method described in detail in Ref.~\cite{Adam:2014qja}. Events are selected according 
to the energy deposited at very large rapidity in the ZN positioned in the Pb-going direction, which mainly detects slow neutrons emitted by the Pb-nucleus as the result of the interaction.
Their emission, according to results obtained in the analysis of lower energy proton-nucleus experiments, is expected to be monotonically related to $N_{\rm coll}$~\cite{Sikler:2003ef}. A centrality selection based on the ZN energy is found to be less biased than other centrality estimators, based on the charged particle multiplicity measurements at central (SPD) or forward (V0) pseudorapidity~\cite{Adam:2014qja}. 
The average number of nucleon-nucleon collisions $\langle N_{\rm coll}\rangle$ for each ZN-selected centrality class is then obtained  by assuming that the charged particle multiplicity measured at central rapidity is proportional to the number of participants $N_{\rm part} = N_{\rm coll} + 1$~\cite{Chemakin:1999jd}.
%In order to compute the nuclear modification factor, the average nuclear overlap function $\langle T^{i}_{\rm pPb}\rangle= \langle N_{\rm coll}^i\rangle /\sigma_{\rm NN}$, where $\langle N_{\rm coll}^i\rangle$ is calculated for each centrality class $i$, and $\sigma_{\rm NN}$ is the nucleon-nucleon inelastic cross section,. The uncertainty on $\langle T^{i}_{\rm pPb}\rangle$ has a constant term (3.4\%), related to the choice of the parameters of the Glauber model~\cite{Miller:2007ri}, plus a centrality-dependent term (1.9\% -- 6.7\%) evaluated using alternative hypotheses when linking $\langle N_{\rm coll}\rangle$ with measured quantities. 
The values of $\langle N_{\rm coll}\rangle$, used in this analysis, are reported in Tab.~\ref{tab:ncoll}, together with their uncertainties. 

\begin{table}[h]
\centering
%\begin{tabular}{c|c|c}
\begin{tabular}{c|c}
%\hline
%ZN centrality class& $\langle T_{\rm pPb}\rangle$  & $\langle N_{\rm coll}\rangle$ \\ \hline
%2--20\% & 0.161 $\pm$ 0.009 $\pm$ 0.005  & 11.3 $\pm$ 0.6 $\pm$ 0.9\\
%20--40\% & 0.136 $\pm$ 0.003 $\pm$ 0.005  & 9.6 $\pm$ 0.2 $\pm$ 0.8\\
%40--60\% & 0.101 $\pm$ 0.005 $\pm$ 0.003  & 7.1 $\pm$ 0.3 $\pm$ 0.6\\
%60--80\% & 0.061 $\pm$ 0.004 $\pm$ 0.002  & 4.3 $\pm$ 0.3 $\pm$ 0.3\\
%80--100\% & 0.030 $\pm$ 0.001 $\pm$ 0.001  & 2.1 $\pm$ 0.1 $\pm$ 0.2\\
\hline
ZN centrality class & $\langle N_{\rm coll}\rangle$ \\ \hline
2--20\% & 11.3 $\pm$ 0.6 $\pm$ 0.9\\
20--40\% & 9.6 $\pm$ 0.2 $\pm$ 0.8\\
40--60\% & 7.1 $\pm$ 0.3 $\pm$ 0.6\\
60--80\% & 4.3 $\pm$ 0.3 $\pm$ 0.3\\
80--100\% & 2.1 $\pm$ 0.1 $\pm$ 0.2\\
\hline
\end{tabular}
\caption{Average numbers of binary nucleon-nucleon collisions, $N_{\rm coll}$, evaluated in the ZN centrality classes used in this analysis. The first quoted systematic uncertainty is uncorrelated, while the second is global.}
\label{tab:ncoll}
\end{table}
 
The centrality classes used in this analysis correspond to 2--20\%, 20--40\%, 40--60\%, 60--80\% and 80--100\% of the measured cross section corresponding to the MB trigger. Very central events (0--2\%) are discarded from the event sample due to a large contamination from pile-up interactions.

\begin{figure}[hbtp]
\centering
\resizebox{1\textwidth}{!}
{\includegraphics{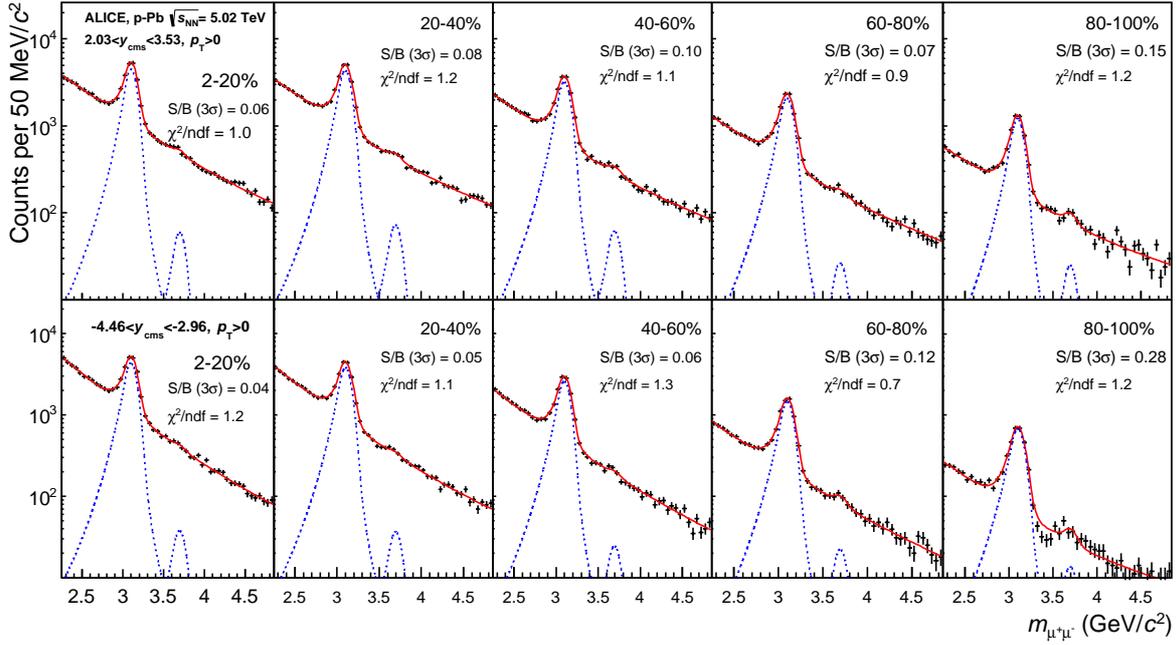}}
\caption{Opposite-sign dimuon invariant mass spectra in ZN centrality classes at forward (top) and backward (bottom) rapidities. The fit curves shown in red in the figure correspond to the sum of signal and background shapes, the former being also shown separately in blue.}
\label{fig:1}
\end{figure}

The estimate of the \psip\ signal is based on binned likelihood fits to the dimuon invariant mass spectra $m_{\mu\mu}$ corresponding to events in the centrality ranges defined above. Details on the procedure, on the fitting functions and on the estimate of systematic uncertainties are discussed in~\cite{Abelev:2014zpa}. The function used in the fit is the sum of a continuum background, mainly related to uncorrelated decays from pions and kaons and to semi-leptonic decays of pairs of hadrons with open heavy flavor, and of resonance shapes corresponding to the \jpsi\ and \psip\ mesons. The background is parameterized by various empirical shapes, directly fitted to the data. The resonances are described by either a Crystal Ball function or a pseudo-gaussian with a mass-dependent width~\cite{Alice:CB}. The main parameters of the \jpsi\ line shapes, i.e.\ mass position and width, are left as free parameters, while the non-gaussian tail parameters are fixed to Monte-Carlo (MC) estimates. The \psip\ line shape parameters, given the less favourable signal over 
background, are fixed relative to those of the \jpsi, assuming that the mass difference and the widths scale according to the MC result.
The results of the fits are shown in Fig.~\ref{fig:1}.

The quality of the fits
is good, with $\chi^2/{\rm ndf}$ ranging from 0.7 to 1.3. The \psip\ signal is 
visible in all the centrality bins, and the signal over background ratio increases from central (0.06 for \mbox{p-Pb} and 0.04 for \mbox{Pb-p}) to peripheral events (0.15 and 0.28, respectively). The number of reconstructed \psip\ for the various centrality bins, $N^{i}_{\psi(2S)\rightarrow\mu^+\mu^-}$, ranges  from $265\pm 73\pm 32$ ($i$= 2--20\%) to $100\pm 29\pm 9$ ($i$= 80--100\%) in \mbox{p-Pb}, where the first uncertainty is statistical and the second one is systematic. The corresponding values for \mbox{Pb-p} are $141\pm 64\pm 13$ ($i$= 2--20\%) and 
$65\pm 20\pm 7$ ($i$= 80--100\%). The systematic uncertainties on the signal extraction are given by the root mean square of the number of \psip\ obtained in  72 fits corresponding to various fitting functions for background and signal, to different fitting ranges, to variations of the non-gaussian tails of the resonance shape, and of the \psip\ mass resolution values. In \mbox{p-Pb}, the systematic uncertainties range between 11 and 13\% from peripheral to central events (11--21\% for \mbox{Pb-p}).
%The \jpsi\ peak is, as expected, much more visible in the invariant mass %spectrum, due to the larger production cross section and branching ratio to %dimuons.

The product of acceptance times efficiency $A\times\epsilon$ for the 
\psip\ resonance was calculated with the MC-based procedure described in Refs.~\cite{Abelev:2013yxa,Abelev:2014zpa}. The values 
are the same as quoted there for the centrality integrated production ($0.270\pm0.014$ for \mbox{p-Pb} and $0.184\pm0.013$ for \mbox{Pb-p}), since it was verified that the tracking efficiency  does not
depend on the centrality of the collision~\cite{Adam:2015jsa}. The quoted errors are the quadratic sum of the systematic uncertainties on tracking, trigger and matching efficiencies and on the choice of the \psip\ $p_{\rm T}$ and $y$ input shapes used in the MC simulations.

The normalization of the \psip\ yield  was calculated according to the procedure described in Ref.~\cite{Adam:2015jsa}. It is based on the evaluation, for each centrality class, of the number of minimum bias events as $N^i_{\rm MB}=F^i_{2\mu/{\rm MB}}\cdot N^i_{2\mu}$, where $N^i_{2\mu}$ is the number of dimuon-triggered events and $F^i_{2\mu/{\rm MB}}$ is the inverse of the probability of having a dimuon triggered in a MB event for that class. The $F^i_{2\mu/{\rm MB}}$-values increase from central to peripheral events and  are $287\pm 3$ and $694\pm 8$ for the 2--20\% centrality class in \mbox{p-Pb} and \mbox{Pb-p} respectively. The corresponding values for the 80--100\% class are $3291\pm 36$ and $3338\pm 35$. 
The systematic uncertainties quoted above (statistical uncertainties are negligible) come from the comparison obtained with two slightly different approaches in the calculation of $F^i_{2\mu/{\rm MB}}$, as detailed in~\cite{Adam:2015jsa}.

In the evaluation of the systematic uncertainties on $F^i_{2\mu/{\rm MB}}$, the presence of interaction pile-up was considered. Pile-up  
can lead to a bias in the evaluation of the centrality of the collision since, for example, the superposition of the signals from two peripheral events in the ZN can fake a more central event. The contribution of pile-up was calculated by detecting
events with multiple interaction vertices in the SPD, and checking via a  Monte-Carlo that the ZN energy distribution can be reproduced assuming a pile-up probability corresponding to the observed interaction rate.
Events in the 0--2\% centrality interval were rejected, as the pile-up contribution becomes significant ($\sim$30\%) in that region.
The effect is small but not negligible in the 2--20\% range, where it amounts to 2.1\% (2.6\%) for \mbox{p-Pb} (\mbox{Pb-p}), and becomes $<1$\% going towards more peripheral events. 
%It has been accounted for by including it in the systematic uncertainty on $F^{\rm i}$.

From the quantities described above, the inclusive cross section for \psip\ production in the centrality bin $i$, times its branching ratio to dimuons ${\rm B.R.}_{\psi(\rm 2S)\rightarrow\mu\mu}$, was calculated with the following expression
\begin{equation}
{\rm B.R.}_{\psi(\rm 2S)\rightarrow\mu^+\mu^-}\sigma^{i, \rm \psi(\rm 2S)}_{\rm pPb}=\frac{N^{i}_{\psi(2S)\rightarrow\mu^+\mu^-}}{(A\times\epsilon)\cdot N_{\rm MB}^{i}}\times \sigma_{\rm MB}
\end{equation}

The ratio $N_{\rm MB}/\sigma_{\rm MB}$, where $N_{\rm MB}$ is the total number of minimum bias events and $\sigma_{\rm MB}$ is the cross section for events satisfying the minimum bias trigger condition, gives the integrated luminosity $L_{\rm int}$. The $\sigma_{\rm MB}$ values were evaluated through a van der Meer scan which gives  $\sigma_{\rm MB}^{\rm pPb}=2.09\pm 0.07$ b and $\sigma_{\rm MB}^{\rm Pbp}=2.12\pm 0.07$ b~\cite{Abelev:2014epa}. 
A determination of the luminosity which makes use of a different reference process, based on the signals released in a \v{C}erenkov counter~\cite{Aamodt:2008zz}, gives a result compatible within 1\%~\cite{Abelev:2014epa}. Therefore, an
additional 1\% uncertainty is added to the $\sigma_{\rm MB}$ values used in the \psip\ cross section determination.

The comparison of the \psip\ and \jpsi\ production cross sections can be performed by calculating the ratio ${\rm B.R.}_{\psi(\rm 2S)\rightarrow\mu^+\mu^-}\sigma_{\psi(\rm 2S)}/{\rm B.R.}_{{\rm J}/\psi\rightarrow\mu^+\mu^-}\sigma_{{\rm J}/\psi}$. In this way, the uncertainties related to the cross section normalization and to the reconstruction efficiency cancel out. The \jpsi\ cross section values that enter this ratio are those reported in~\cite{Adam:2015jsa}, with the value for the centrality interval 2--20\% obtained by summing the 2--10\% and 10--20\% results. This ratio can be further normalized to the corresponding measurement in pp collisions. This quantity, called double ratio in the following, gives direct access to modifications in the \psip\ production relative to that of the \jpsi, going from \mbox{pp} to \mbox{p-Pb} collisions. Due to the lack of precise pp data at $\sqrt{s}=5.02$~TeV, the results obtained at $\sqrt{s}=7$~TeV~\cite{Abelev:2014qha} were used instead. This choice is justified from the fact that the $\sqrt{s}$- and $y$-dependence of the cross section ratio is known to be weak in the TeV beam energy range. An 8\% systematic uncertainty has been included, corresponding to the maximum estimated size of the variation of the ratio between the two energies~\cite{Abelev:2014zpa}.

The estimate of the nuclear modification factors $Q^{i,\psi(2S)}_{\rm pPb}$ as a function of centrality is performed as the product of the corresponding $Q^{i,{\rm J}/\psi}_{\rm pPb}$ for the \jpsi~\cite{Adam:2015jsa} (except for the 2--20\% centrality interval where $Q^{i,{\rm J}/\psi}_{\rm pPb}$ was re-computed by merging the 2--10\% and 10--20\% bins) and the double ratio between the \psip\ and \jpsi\ cross sections in \mbox{p-Pb} and pp collisions:

\begin{equation}
Q^{i,\psi(\rm 2S)}_{\rm pPb}=Q^{i,{\rm J}/\psi}_{\rm pPb}\cdot \frac{\sigma^{i,\psi(\rm 2S)}_{\rm pPb}}
{\sigma^{i,{\rm J}/\psi}_{\rm pPb}}\cdot 
\frac{\sigma^{{\rm J}/\psi}_{\rm pp}}{\sigma^{\psi(\rm 2S)}_{\rm pp}}
\label{eq:2}
\end{equation}

The uncertainties are obtained combining those on $Q^{i,{\rm J}/\psi}_{\rm pPb}$~\cite{Adam:2015jsa} with those on the double ratio, avoiding a double counting of the \jpsi\ related uncertainties. The notation $Q^{i,\psi(2S)}_{\rm pPb}$, rather than the more usual $R^{i,\psi(2S)}_{\rm pPb}$, is used in this Letter, 
to draw attention to possible residual biases in the centrality determination, related to the loose correlation between the centrality estimators and the corresponding collision geometry~\cite{Adam:2014qja}.

Table~\ref{tab:syst} summarizes the values of the systematic uncertainties on the various ingredients that enter the cross section determination and the calculation of the nuclear modification factor.

\begin{table}[h]
\centering
\begin{tabular}{l|c|c}
\hline
Source of uncertainty& $\sigma^{\psi(\rm 2S)}_{\rm pPb}$, $Q^{\psi(\rm 2S)}_{\rm pPb}$  & $\sigma^{\psi(\rm 2S)}_{\rm Pbp}$, $Q^{\psi(\rm 2S)}_{\rm Pbp}$ \\ 
 &  2.03$<y_{\rm cms}<$3.53 &  -4.46$<y_{\rm cms}<$-2.96  \\ \hline
Tracking efficiency (I) &  4 & 6\\
Trigger efficiency (I) &  3 & 3.4\\
Matching efficiency (I)&  1 & 1 \\ 
Signal extraction &  10.8 $-$ 13.4 & 10.8 $-$ 20.9\\
MC input &  1.8 & 2.5\\
$\sigma_{\rm MB}$ (I) &  3.3 & 3.0\\
$\sigma_{\rm MB}$ (I,II) &  1.6 & 1.6\\
\hline
\end{tabular}
\caption{Systematic uncertainties, in percentage, on the \psip\ cross sections and nuclear modification factors. 
For centrality-dependent quantities, the range of variation is given. 
Type I uncertainties are correlated over centrality, while type II are correlated between the forward and the backward rapidity regions. When no indication is given, the uncertainties are uncorrelated.
The uncertainty on $\sigma_{\rm MB}$ is related to the \psip\ cross section only.}
\label{tab:syst}
\end{table}

\section{Results}

\begin{figure}[hbtp]
\centering
\resizebox{0.49\textwidth}{!}
{\includegraphics{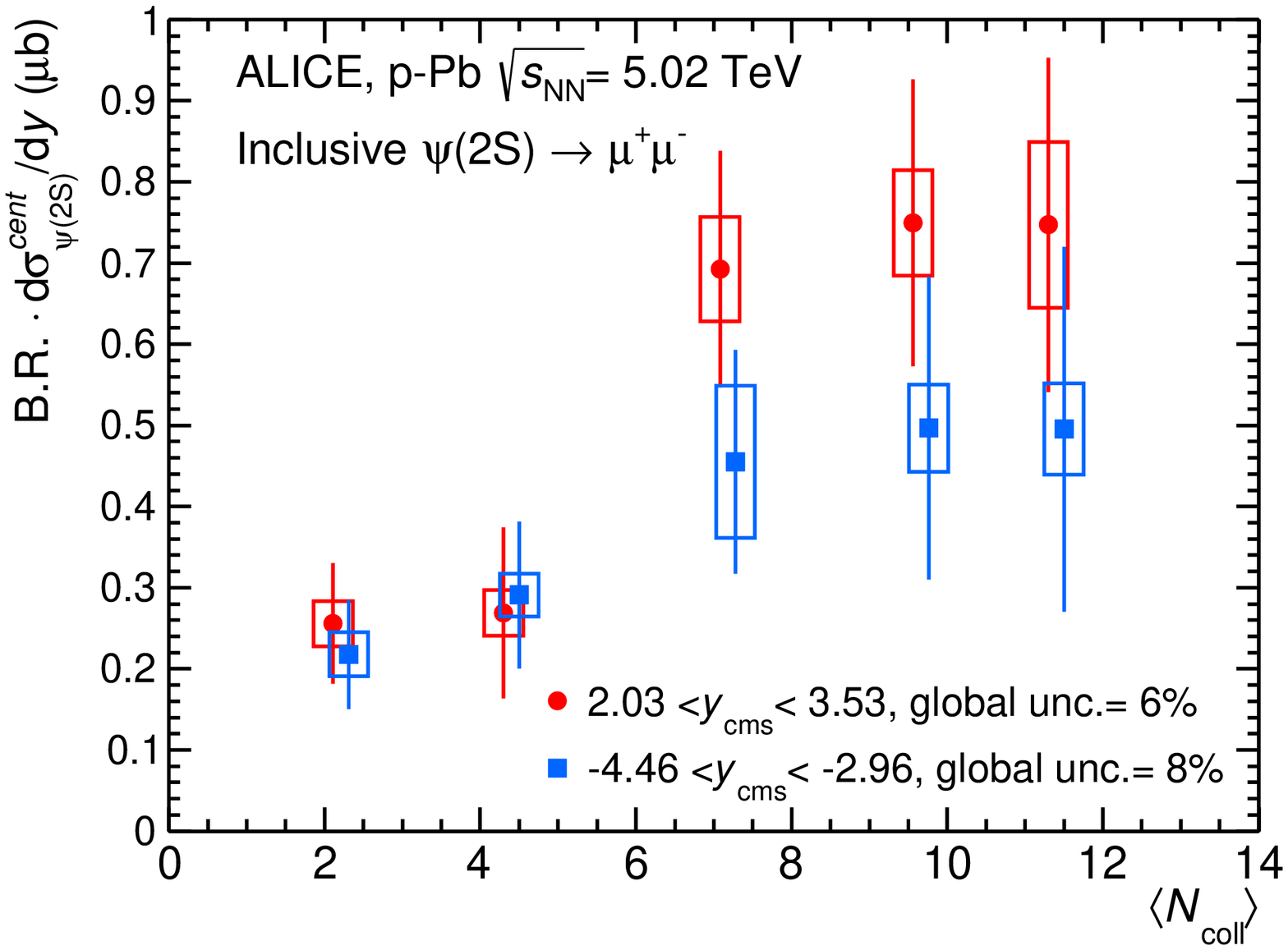}}
\resizebox{0.49\textwidth}{!}
{\includegraphics{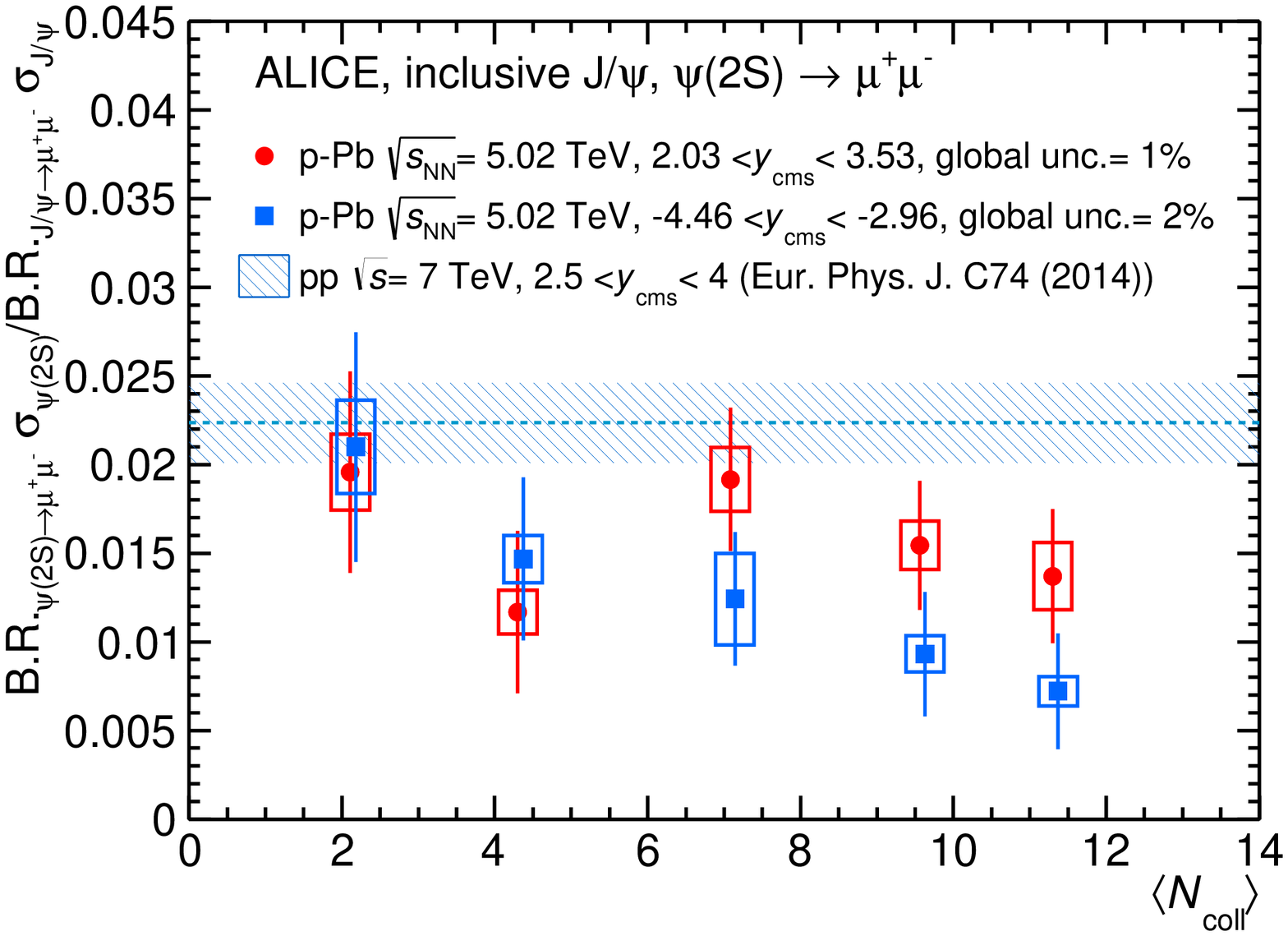}}
\caption{Left: \psip\ production cross sections shown as a function of $\langle N_{\rm coll}\rangle$  for both \mbox{p-Pb} and \mbox{Pb-p} collisions. 
Right: ${\rm B.R.}_{\psi(\rm 2S)\rightarrow\mu^+\mu^-}\sigma_{\psi(\rm 2S)}/{\rm B.R.}_{{\rm J}/\psi\rightarrow\mu^+\mu^-}\sigma_{{\rm J}/\psi}$ shown as a function of 
$\langle N_{\rm coll}\rangle$ and compared to the pp value (line), with a band representing its uncertainty.
In both figures, vertical error bars correspond to statistical uncertainties, while the open boxes represent the systematic uncertainties. The \mbox{Pb-p} points are slightly shifted in $\langle N_{\rm coll}\rangle$ to improve visibility.
}
\label{fig:2}
\end{figure}

The \psip\ production cross sections as a function of the centrality of the collision, expressed via $\langle N_{\rm coll}\rangle$, are plotted in Fig.~\ref{fig:2} (left). 
As expected, their values increase with $\langle N_{\rm coll}\rangle$. 
%Within uncertainties, the cross section values at forward and backward rapidity are compatible.

In Fig.~\ref{fig:2} (right) the ratio ${\rm B.R.}_{\psi(\rm 2S)\rightarrow\mu^+\mu^-}\sigma_{\psi(\rm 2S)}/{\rm B.R.}_{{\rm J}/\psi\rightarrow\mu^+\mu^-}\sigma_{{\rm J}/\psi}$ is shown as a function of 
$\langle N_{\rm coll}\rangle$ and compared with the corresponding value for pp collisions. Despite the large uncertainties, the data
suggest a decreasing trend from peripheral to central events, in particular at backward rapidity, indicating a suppression of the \psip\ production relative to the \jpsi. While for peripheral collisions the cross section ratios are consistent with the pp value, they become a factor 2--3 smaller for central events, in both rapidity ranges. As remarked in Sect.~\ref{sec:3}, the pp cross section ratio measured at $\sqrt{s}=7$~TeV has been used, including an 8\% additional uncertainty to account for its possible $\sqrt{s}$- and $y$-dependence. 

\begin{figure}[hbtp]
\centering
\resizebox{0.65\textwidth}{!}
{\includegraphics{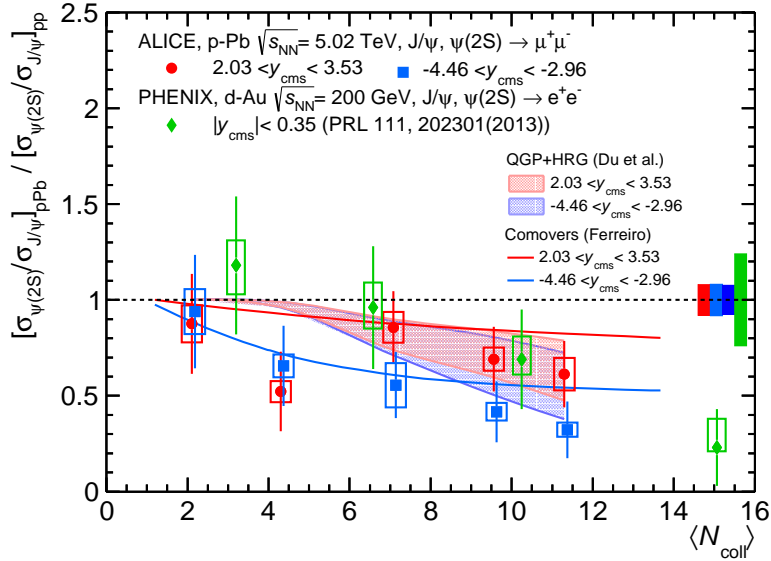}}
%{\includegraphics{Figures/DoubleRatio}}
\caption{Double ratio $[\sigma_{\psi(\rm 2S)}/\sigma_{{\rm J}/\psi}]_{\rm pPb}/[\sigma_{\psi(\rm 2S)}/\sigma_{{\rm J}/\psi}]_{\rm pp}$ for \mbox{p-Pb} and \mbox{Pb-p} collisions, shown as a function of $\langle N_{\rm coll}\rangle$ (\mbox{Pb-p} points are slightly shifted in $\langle N_{\rm coll}\rangle$ to improve visibility). The data are compared to PHENIX mid-rapidity results~\cite{Adare:2013ezl} and to the theoretical calculations of Ref.~\cite{Ferreiro:2014bia} and~\cite{Du:2015wha}. The boxes around unity correspond to the global systematic uncertainties at forward (red box) and backward (blue box) rapidities. 
The grey box is a global systematic uncertainty common to both  \mbox{p-Pb} rapidity ranges, while the green box refers to the PHENIX results.}
\label{fig:4}
\end{figure}

The degree of suppression of \psip\ is directly quantified in Fig.~\ref{fig:4} where the double ratio between the
\psip\ and \jpsi\ cross sections in \mbox{p-Pb} and pp collisions is shown. The result is compared with two theoretical calculations. The first is based on a scenario where the resonances may be dissociated via interactions with the partons or hadrons produced in the collision in the same rapidity region (co-movers)~\cite{Ferreiro:2014bia}.
The model includes contributions from nuclear shadowing, based on the EPS09 LO parameterization~\cite{Eskola:2009uj}, and a co-mover interaction term, with dissociation cross sections $\sigma^{\rm co-J/\psi}=0.65$~mb and  $\sigma^{\rm co-\psi(2S)}=6$~mb, these values being fixed from fits to low-energy experimental data~\cite{Armesto:1997sa}. The effect of co-movers is larger at backward rapidity since their density is larger in that region. The calculated co-mover densities are compatible with the measured experimental charged particle multiplicities~\cite{ALICE:2012xs}.
The calculation reproduces well the measured values of the double ratio. 
Shadowing effects are very similar for the two mesons and in this model they are assumed to cancel out in the double ratio, so that only co-mover absorption plays a role. 
The second model (QGP+HRG) is based on a thermal-rate equation framework~\cite{Zhao:2010nk} which also implements the dissociation of charmonia in a hadron resonance gas, including a total of 52 non-strange and single-strange meson species, up to a mass of 2 GeV/$c^2$~\cite{Du:2015wha}. The fireball evolution includes the transition from a short QGP phase into the hadron resonance gas, through a mixed phase. 
The shadowing effects, implemented through the EPS09 parametrization, cancel out in the double ratio, as in the previous model. The result of the calculation, also shown in Fig.~\ref{fig:4}, is in fair agreement with the measured values, in particular for central collisions. The model uncertainties are dominated by the evaluation of the charmonium dissociation rates.
The ALICE result is also compared to mid-rapidity ($|y|<0.35$) PHENIX data~\cite{Adare:2013ezl} in Fig.~\ref{fig:4}. Remarkably, in spite of the very different $\sqrt{s_{\rm NN}}$ and $y_{\rm cms}$ values, the observed patterns as a function of centrality are similar. It should also be noted that the PHENIX result can be qualitatively described in a hadronic dissociation scenario, as discussed in ~\cite{Ferreiro:2014bia,Du:2015wha}.

\begin{figure}[hbtp]
\centering
\resizebox{0.49\textwidth}{!}
{\includegraphics{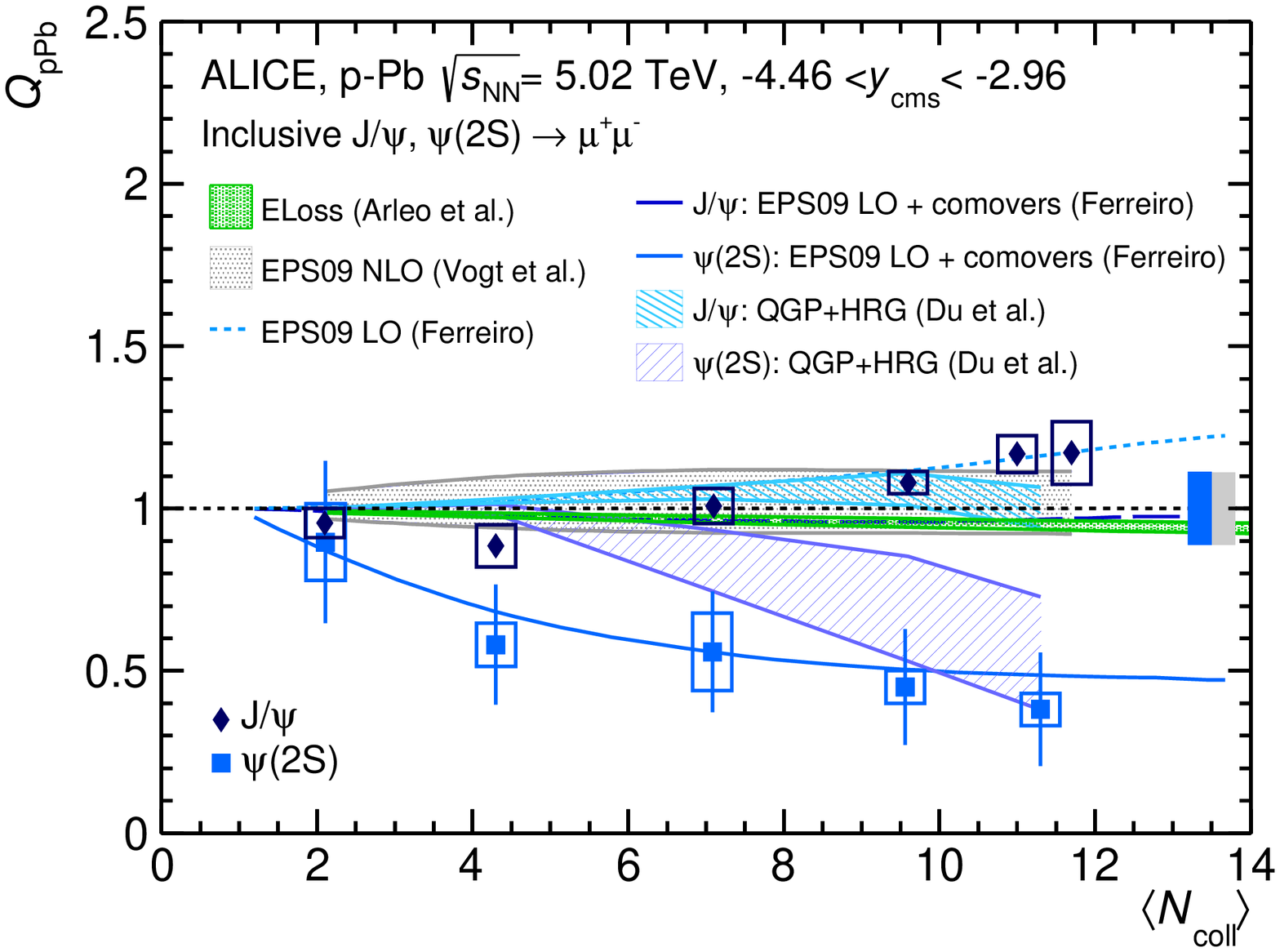}}
\resizebox{0.49\textwidth}{!}
{\includegraphics{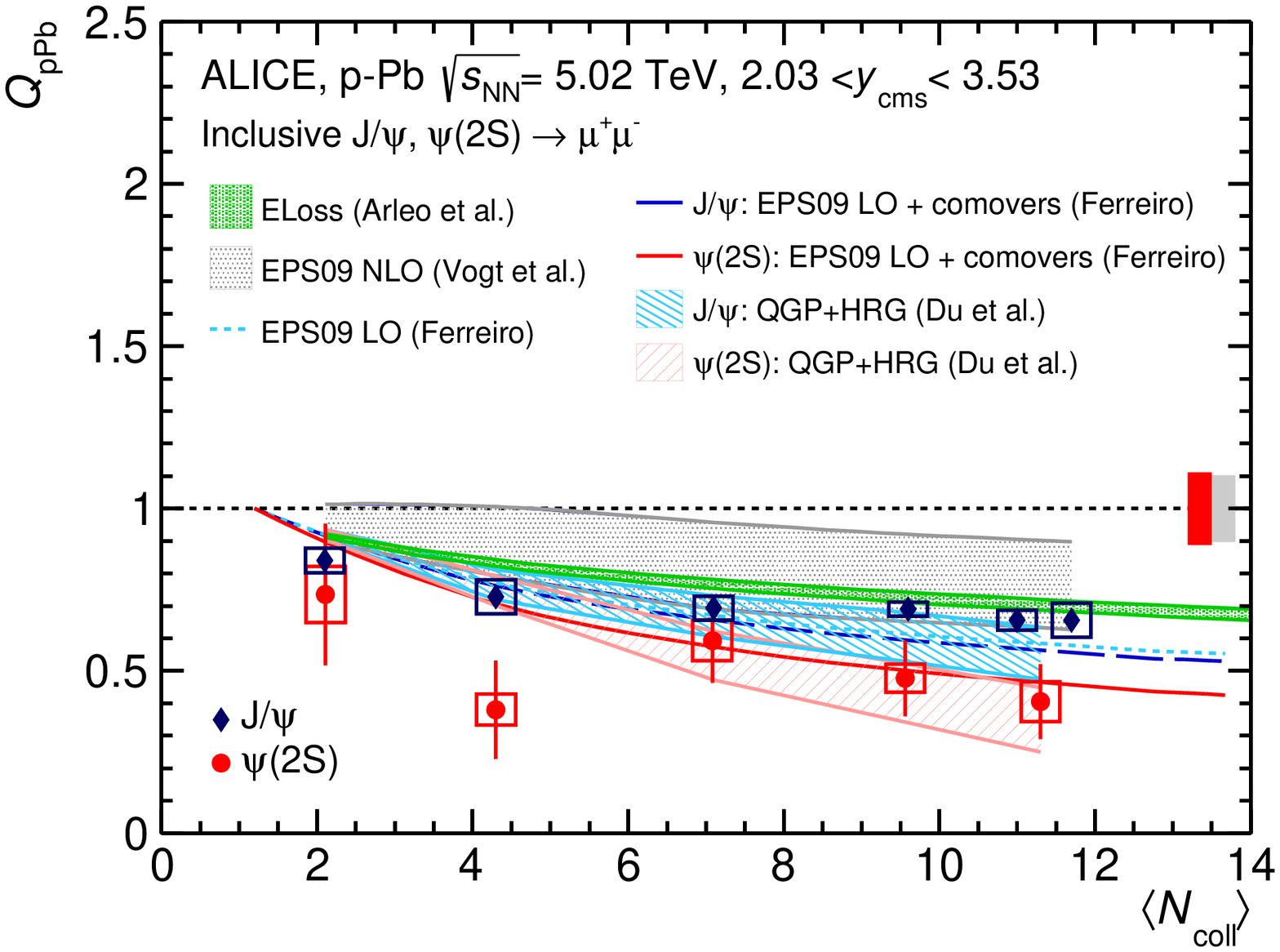}}
\caption{\jpsi~\cite{Adam:2015jsa} and \psip\ nuclear modification factors, $Q_{\rm pPb}$, shown as a function of $\langle N_{\rm coll}\rangle$ for the backward (left) and forward (right) rapidity regions and compared to theoretical models~\cite{Vogt:2015uba,Ferreiro:2014bia,Du:2015wha,Arleo:2013zua}. The boxes around unity correspond to the global \psip\  systematic uncertainties at forward (red box) and backward (blue box) rapidities. 
The grey box is a global systematic uncertainty common to both \jpsi\ and \psip.}
\label{fig:5}
\end{figure}

In Fig.~\ref{fig:5} the nuclear modification factor for \psip\ mesons is shown as a function of centrality, separately for forward and backward rapidities. 
In both regions, a trend towards an increasing suppression can be seen when moving from peripheral to central collisions. The corresponding $Q^{{\rm J}/\psi}_{\rm pPb}$ values~\cite{Adam:2015jsa} are also shown. 
At forward-$y$ there is an indication for a smaller $Q^{\psi(2S)}_{\rm pPb}$ with respect to $Q^{{\rm J}/\psi}_{\rm pPb}$. The difference between the \psip\ and the \jpsi\ nuclear modification factors amounts, for central events, to 1.9$\sigma$, while, integrating over centrality, the corresponding quantity is 2.3$\sigma$. At backward-$y$ the suppression patterns for the \jpsi\ and the \psip\ are different, with $Q^{{\rm J}/\psi}_{\rm pPb}\sim 1$ (or even slightly larger), and a strong suppression for the \psip. In the most central collisions, the difference between the measured $Q_{\rm pPb}$ corresponds to 4.3$\sigma$, while, integrating over centrality, suppressions differ by 4.1$\sigma$. 
The results are compared to calculations including either only shadowing (EPS09 LO~\cite{Ferreiro:2014bia}, EPS09 NLO~\cite{Vogt:2015uba}) or only coherent energy loss~\cite{Arleo:2013zua} and to models implementing final state interactions (co-movers~\cite{Ferreiro:2014bia}, QGP+HRG~\cite{Du:2015wha}). While the \jpsi\ results are reproduced by shadowing/energy loss calculations, additional final state effects, as those discussed in the context of Fig.~\ref{fig:4},  are needed to describe the \psip\ results, in particular at backward rapidity.

\begin{figure}[hbtp]
\centering
\resizebox{0.65\textwidth}{!}
{\includegraphics{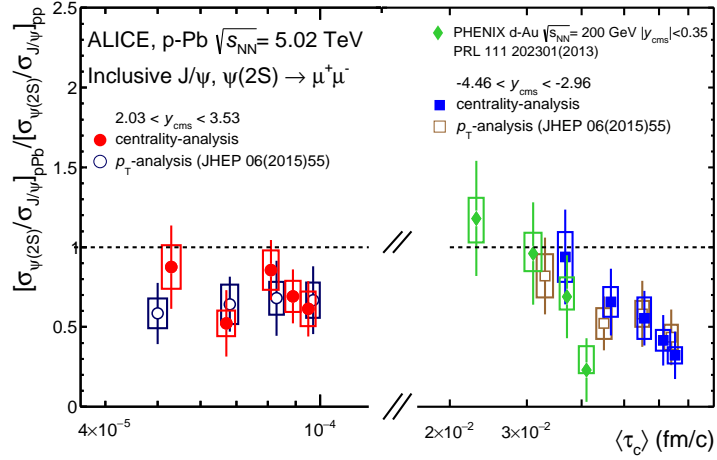}}
\caption{Double ratio $[\sigma_{\psi(\rm 2S)}/\sigma_{{\rm J}/\psi}]_{\rm pPb}/[\sigma_{\psi(\rm 2S)}/\sigma_{{\rm J}/\psi}]_{\rm pp}$ shown as a function of $\tau_c$ for the backward and forward rapidity regions. For each $y$-range, the two sets of points were obtained from the centrality analysis and from the $p_{\rm T}$-dependent analysis of Ref.~\cite{Abelev:2014zpa}. Statistical uncertainties are shown as lines, while the total systematic uncertainties are shown as boxes around the points. The results of a corresponding analysis carried out on the PHENIX mid-rapidity data~\cite{Adare:2013ezl} is also shown. The box around unity represents the PHENIX global systematic uncertainty. For the ALICE results, the global uncertainties are different for the various data sets, and are included in the boxes around the points}
\label{fig:6}
\end{figure}

Finally, the double ratios are shown in Fig.~\ref{fig:6} as a function of the  pair crossing time $\tau_{\rm c}$ in nuclear matter~\cite{McGlinchey:2012bp}. This quantity can be calculated as $\tau_{\rm c}= \langle L \rangle/(\beta_{\rm z}\gamma)$ where $\langle L \rangle$ is the average thickness of nuclear matter crossed by the pair, which was evaluated, for each centrality class, using the Glauber model~\cite{Miller:2007ri}, \hbox{$\beta_{\rm z}=\tanh y_{c\overline c}^{\rm rest}$} is the velocity of the $c\overline c$ along the beam direction in the nucleus rest frame,  $\gamma=E_{c\overline c}/m_{c\overline c}$ and \hbox{$E_{c\overline c}=m_{\rm T,c\overline c}\cosh y_{c\overline c}^{\rm rest}$}. The value $m_{c\overline c}=3.4$ GeV/$c^2$ was chosen for the (average) mass of the evolving $c{\overline c}$ pair~\cite{McGlinchey:2012bp, Arleo:1999af}, while $m_{\rm T,c\overline c}$ was calculated in each centrality bin starting from the measured J/$\psi$ $\langle p_{\rm T}\rangle$ values~\cite{Adam:2015jsa}. We use the J/$\psi$ $\langle p_{\rm T}\rangle$  as a proxy for the average $p_{\rm T}$ of the $c{\overline c}$  pair, as the $\psi(2S)$ statistics is too low to extract a corresponding $\langle p_{\rm T}\rangle$ value. If we assume instead that $\langle p_{\rm T}^{\psi(2S)}\rangle \sim 1.1 \langle p_{\rm T}^{{\rm J}/\psi}\rangle$ as measured by LHCb in pp collisions at $\sqrt{s}=7$ TeV~\cite{Aaij:2012ag,Aaij:2013nlm}, the $\tau_{\rm c}$ values would decrease by $\sim 4$\%. Other sources of uncertainties on $\tau_{\rm c}$ include the uncertainties on the measured J/$\psi$ $\langle p_{\rm T}\rangle$, which contribute less than 1\%, and those on $\langle L \rangle$, which are dominant and of the order of 10\%.
In Fig.~\ref{fig:6} we show the double ratio as a function of $\tau_{\rm c}$ in the two rapidity regions. Different $\tau_{\rm c}$ intervals can also be selected by slicing the events in bins of $p_{\rm T}$ (see Ref.~\cite{Adam:2015iga}), varying, in this way, the  $\gamma$ values of the $c\overline c$.  
The double ratio results, obtained in~\cite{Adam:2015iga}, are therefore also shown in Fig.~\ref{fig:6} at their corresponding average $\tau_{\rm c}$ values.
In the double ratio one effectively removes, as discussed above, initial state effects, so that Fig.~\ref{fig:6} shows the $\tau_{\rm c}$ dependence of final state effects on \psip\ compared to \jpsi. The two sets of results,  corresponding to a slicing of the events in centrality or in $p_{\rm T}$, are in good agreement. 
At backward-$y$, where the largest $\tau_{\rm c}$ values are reached, a clearly decreasing trend can be observed. The average resonance  formation time $\tau_{\rm f}$ is, according to most theory estimates~\cite{McGlinchey:2012bp,Hufner:2000jb,Kharzeev:1999bh}, larger by at least a factor $\sim 2$ than the accessible $\tau_{\rm c}$ range. On the other hand, the width of the $\tau_{\rm f}$ distribution is expected to be non-negligible~\cite{Kharzeev:1999bh}, and it cannot be excluded that at least a fraction of the $c\overline c$ pairs hadronizes inside the nucleus. 
Therefore, the observed behaviour is likely due to a combination of final state effects which take place outside the nucleus, as e.g.\ interaction with a hadronic resonance gas, and dissociation effects on the fully formed resonance, due to nuclear matter, and taking place inside the nucleus. The relative importance of the two mechanisms is difficult to quantify in such a simple analysis and  quantitative theoretical studies, also exploring alternative mechanisms, are needed. At forward rapidity, where $\tau_{\rm c}$ becomes smaller than $\tau_{\rm f}$ by about 2--3 orders of magnitude, the interaction with nuclear matter is not expected to play any significant role. 
The results of a similar analysis carried out on PHENIX mid-rapidity data~\cite{Adare:2013ezl} are also shown in Fig.~\ref{fig:6}. Within uncertainties, a scaling of the ALICE and PHENIX double ratio values with $\tau_{\rm c}$ is observed.

\section{Conclusions}

The centrality dependence of the \psip\ production in \mbox{p-Pb} collisions at $\sqrt{s_{\rm NN}} = 5.02$~TeV was  measured in five intervals, using the ZN energy as an estimator. The ratio ${\rm B.R.}_{\psi(\rm 2S)\rightarrow\mu^+\mu^-}\sigma_{\psi(\rm 2S)}/{\rm B.R.}_{{\rm J}/\psi\rightarrow\mu^+\mu^-}\sigma_{{\rm J}/\psi}$ is compatible with pp measurements in peripheral events, whereas a decrease is observed towards central events, showing that the \psip\ state is suppressed with respect to the \jpsi\ state. The results on the nuclear modification factors, together with the corresponding model comparisons, show that effects such as shadowing or energy loss are enough to explain the \jpsi\ behaviour, while additional mechanisms are needed to describe the \psip\ suppression.
Theoretical models that include final state interactions are able to reproduce such a suppression.  
A study of the double ratio $[\sigma_{\psi(\rm 2S)}/\sigma_{{\rm J}/\psi}]_{\rm pPb}/[\sigma_{\psi(\rm 2S)}/\sigma_{{\rm J}/\psi}]_{\rm pp}$, as a function of the crossing time $\tau_{\rm c}$, shows that at forward-$y$ the $\tau_{\rm c}$ values are much shorter than the resonance formation time $\tau_{\rm f}$, excluding any significant role of final state interactions with nuclear matter.  Effects occurring at later times, such as the break-up by co-movers in the hadronic gas, are suitable candidates for an explanation of the observed \psip\ suppression. 
%The same holds at backward-$y$ where the $\tau_{\rm c}$ values, although significantly larger, are still smaller than 
%$\tau_{\rm f}$.
At backward-$y$ the $\tau_{\rm c}$ values, although significantly larger, are still smaller than $\tau_{\rm f}$. However, the observed scaling of the double ratios with $\tau_{\rm c}$ may be suggestive of an effect at least partly related to a dissociation of the fully-formed resonance in nuclear matter.
               %%%%%%%%%%% put the body of the article here
% 
%%%%%%%% acknowledgements
\newenvironment{acknowledgement}{\relax}{\relax}
\begin{acknowledgement}
\section*{Acknowledgements}
% $Id: acknowledgements.tex 2287 2015-10-20 20:30:56Z loizides $
% Version: Nov 2015

The ALICE Collaboration would like to thank all its engineers and technicians for their invaluable contributions to the construction of the experiment and the CERN accelerator teams for the outstanding performance of the LHC complex.
The ALICE Collaboration gratefully acknowledges the resources and support provided by all Grid centres and the Worldwide LHC Computing Grid (WLCG) collaboration.
The ALICE Collaboration acknowledges the following funding agencies for their support in building and
running the ALICE detector:
State Committee of Science,  World Federation of Scientists (WFS)
and Swiss Fonds Kidagan, Armenia;
Conselho Nacional de Desenvolvimento Cient\'{\i}fico e Tecnol\'{o}gico (CNPq), Financiadora de Estudos e Projetos (FINEP),
Funda\c{c}\~{a}o de Amparo \`{a} Pesquisa do Estado de S\~{a}o Paulo (FAPESP);
National Natural Science Foundation of China (NSFC), the Chinese Ministry of Education (CMOE)
and the Ministry of Science and Technology of China (MSTC);
Ministry of Education and Youth of the Czech Republic;
Danish Natural Science Research Council, the Carlsberg Foundation and the Danish National Research Foundation;
The European Research Council under the European Community's Seventh Framework Programme;
Helsinki Institute of Physics and the Academy of Finland;
French CNRS-IN2P3, the `Region Pays de Loire', `Region Alsace', `Region Auvergne' and CEA, France;
German Bundesministerium fur Bildung, Wissenschaft, Forschung und Technologie (BMBF) and the Helmholtz Association;
General Secretariat for Research and Technology, Ministry of Development, Greece;
National Research, Development and Innovation Office (NKFIH), Hungary;
Department of Atomic Energy and Department of Science and Technology of the Government of India;
Istituto Nazionale di Fisica Nucleare (INFN) and Centro Fermi -
Museo Storico della Fisica e Centro Studi e Ricerche ``Enrico Fermi'', Italy;
Japan Society for the Promotion of Science (JSPS) KAKENHI and MEXT, Japan;
Joint Institute for Nuclear Research, Dubna;
National Research Foundation of Korea (NRF);
Consejo Nacional de Cienca y Tecnologia (CONACYT), Direccion General de Asuntos del Personal Academico(DGAPA), M\'{e}xico, Amerique Latine Formation academique - 
European Commission~(ALFA-EC) and the EPLANET Program~(European Particle Physics Latin American Network);
Stichting voor Fundamenteel Onderzoek der Materie (FOM) and the Nederlandse Organisatie voor Wetenschappelijk Onderzoek (NWO), Netherlands;
Research Council of Norway (NFR);
National Science Centre, Poland;
Ministry of National Education/Institute for Atomic Physics and National Council of Scientific Research in Higher Education~(CNCSI-UEFISCDI), Romania;
Ministry of Education and Science of Russian Federation, Russian
Academy of Sciences, Russian Federal Agency of Atomic Energy,
Russian Federal Agency for Science and Innovations and The Russian
Foundation for Basic Research;
Ministry of Education of Slovakia;
Department of Science and Technology, South Africa;
Centro de Investigaciones Energeticas, Medioambientales y Tecnologicas (CIEMAT), E-Infrastructure shared between Europe and Latin America (EELA), 
Ministerio de Econom\'{i}a y Competitividad (MINECO) of Spain, Xunta de Galicia (Conseller\'{\i}a de Educaci\'{o}n),
Centro de Aplicaciones Tecnológicas y Desarrollo Nuclear (CEA\-DEN), Cubaenerg\'{\i}a, Cuba, and IAEA (International Atomic Energy Agency);
Swedish Research Council (VR) and Knut $\&$ Alice Wallenberg
Foundation (KAW);
Ukraine Ministry of Education and Science;
United Kingdom Science and Technology Facilities Council (STFC);
The United States Department of Energy, the United States National
Science Foundation, the State of Texas, and the State of Ohio;
Ministry of Science, Education and Sports of Croatia and  Unity through Knowledge Fund, Croatia;
Council of Scientific and Industrial Research (CSIR), New Delhi, India;
Pontificia Universidad Cat\'{o}lica del Per\'{u}.
    %%%%%%% get the lates version before submitting
\end{acknowledgement}

\bibliographystyle{utphys}
\bibliography{psi2S-alicepreprint_070416}
\newpage

%%
%%\newpage
%%
%%\input{}               %%%%%%%%%%% put your appendices here
%%
%%%%%%%%%% appendix with author list
\appendix
\section{The ALICE Collaboration}
\label{app:collab}

\bigskip 

J.~Adam$^{\rm 39}$, 
D.~Adamov\'{a}$^{\rm 84}$, 
M.M.~Aggarwal$^{\rm 88}$, 
G.~Aglieri Rinella$^{\rm 35}$, 
M.~Agnello$^{\rm 110}$, 
N.~Agrawal$^{\rm 47}$, 
Z.~Ahammed$^{\rm 133}$, 
S.~Ahmad$^{\rm 18}$, 
S.U.~Ahn$^{\rm 68}$, 
S.~Aiola$^{\rm 137}$
% ...

\bigskip

\bigskip 

\textbf{\Large Affiliation Notes}

\bigskip 

% ...
$^{\rm IV}$ Also at: M.V. Lomonosov Moscow State University, D.V. Skobeltsyn Institute of Nuclear, Physics, Moscow, Russia\\

\bigskip

\bigskip 

\textbf{\Large Collaboration Institutes}

\bigskip 

% ...
$^{18}$ Department of Physics, Aligarh Muslim University, Aligarh, India\\
% ...
$^{35}$ European Organization for Nuclear Research (CERN), Geneva, Switzerland\\
% ...
$^{39}$ Faculty of Nuclear Sciences and Physical Engineering, Czech Technical University in Prague, Prague, Czech Republic\\
% ...
$^{47}$ Indian Institute of Technology Bombay (IIT), Mumbai, India\\
% ...
$^{68}$ Korea Institute of Science and Technology Information, Daejeon, South Korea\\
% ...
$^{84}$ Nuclear Physics Institute, Academy of Sciences of the Czech Republic, \v{R}e\v{z} u Prahy, Czech Republic\\
% ...
$^{88}$ Physics Department, Panjab University, Chandigarh, India\\
% ...
$^{110}$ Sezione INFN, Turin, Italy\\
% ...
$^{133}$ Variable Energy Cyclotron Centre, Kolkata, India\\
% ...
$^{137}$ Yale University, New Haven, Connecticut, United States\\
% ...

  %%%%%%% get the latest version before submitting
%%
%%\input{}              %%%%%%%%%%%% put the references here
%%
\end{document}